# Electromagnetic enhancement generated by $\hat{\mathbf{A}}\cdot\hat{\mathbf{p}}$ term of cavity quantum electrodynamics demonstrated by single coupled systems between plasmon and molecular exciton

Tamitake Itoh[1]*, Yuko S. Yamamoto[2]

[1]Nano-Bioanalysis Research Group, Health Research Institute, National Institute of Advanced Industrial Science and Technology (AIST), Takamatsu, Kagawa 761-0395, Japan

[2]School of Materials Science, Japan Advanced Institute of Science and Technology (JAIST), Nomi, Ishikawa 923-1292, Japan

*Corresponding author: tamitake-itou@aist.go.jp

# ABSTRACT

In non-relativistic quantum electrodynamics, an electromagnetic (EM) interaction between a photon and a molecular exciton can be expressed by a $\hat{\mathbf{A}} \cdot \hat{\mathbf{p}}$ term and a $\hat{\mathbf{A}}^2$ term, where $\hat{\mathbf{A}}$ and $\hat{\mathbf{p}}$ are the operators of the vector potential of the EM field and the momentum of the exciton, respectively. We developed a method for investigating the contribution of the $\hat{\mathbf{A}} \cdot \hat{\mathbf{p}}$ and $\hat{\mathbf{A}}^2$ terms to EM enhancement, which occurs in coupled systems composed of a plasmon polariton and a molecular exciton. The spectral shapes of the $\hat{\mathbf{A}} \cdot \hat{\mathbf{p}}$ and $\hat{\mathbf{A}}^2$ terms, and the EM enhancement were experimentally obtained from absorption, Rayleigh scattering, and ultrafast surface enhanced fluorescence (ultrafast SEF) of the systems, respectively. The relationships between them reveal that the absorption spectra correctly reproduce EM enhancement, indicating that ultrafast SEF can be described as a two-step process using the $\hat{\mathbf{A}} \cdot \hat{\mathbf{p}}$ terms. Furthermore, we demonstrate that the origin of spectral deviation between Rayleigh scattering and EM enhancement is subradiant plasmon resonance, which spectra are visualized in the absorption not in Rayleigh scattering, with numerical calculation based on electromagnetism.

**I. INTRODUCTION**

The cross-sections of the electromagnetic (EM) interactions of coupled systems composed of a plasmon polariton and a molecular exciton are largely enhanced by a tightly confined EM field inside the plasmonic cavity [1]. In particular, a molecule located inside the cavity, such as a nanogap between metallic nanoparticle (NP) aggregates exhibits an EM enhancement factor of up to $10^{10}$ in Raman scattering, realizing single molecule (SM) spectroscopy under the resonant Raman condition [2-5]. This phenomenon is called surface-enhanced resonant Raman scattering (SERRS) and such a nanogap is called a hotspot (HS). HSs have received considerable attention in the cavity quantum electrodynamics (cQED) field because HSs exhibit various exotic phenomena, such as a nonlinear spectroscopy with CW laser excitation [6], ultra-fast surface-enhanced fluorescence (ultrafast SEF) [7,8], vibrational pumping [9], and the field gradient effect [10]. Furthermore, the EM coupling energy between a plasmon polariton and a molecular exciton at HSs exceeds several hundred meV, resulting in new physics and chemistry, e.g., strong to ultrastrong coupling [11], molecular optomechanics [12], and polariton chemistry [13,14].

In cQED, the EM interaction between a photon and a molecular exciton is expressed by the $\hat{\mathbf{A}} \cdot \hat{\mathbf{p}}$ and $\hat{\mathbf{A}}^2$ terms, where $\hat{\mathbf{A}}$ and $\hat{\mathbf{p}}$ are the operators of the vector potential and momentum, respectively. [15,16]. Under the dipole approximation, in which the light wavelength is enough larger than the molecule size as in Figs. 1(a1) and 1(a2), the $\hat{\mathbf{A}} \cdot \hat{\mathbf{p}}$ and $\hat{\mathbf{A}}^2$ terms generate absorption (or emission) and Rayleigh scattering, respectively, in the first order terms of a perturbation calculation. However, it is not obvious that a dipole approximation is applicable in describing EM interactions inside HSs, in which EM fields are confined within the order of nanometers as in Figs. 1(b1) and 1(b2). Indeed, a breakdown of the dipole approximation has been observed in vibrational spectroscopy using HSs [10,17-19]. Furthermore, strong coupling between molecular excitons and vacuum EM field tightly confined in HSs are reported as illustrated in Fig. 1(b3) [10-14]. Therefore, a method for evaluating the contribution of the $\hat{\mathbf{A}} \cdot \hat{\mathbf{p}}$ and $\hat{\mathbf{A}}^2$ terms to EM enhancement needs to be developed for various plasmonic systems having HSs [1,11].

In this study, the contributions of the $\hat{\mathbf{A}} \cdot \hat{\mathbf{p}}$ and $\hat{\mathbf{A}}^2$ terms to EM enhancement at HSs were evaluated by ultrafast SEF using single silver NP dimer gaps as the HSs. The ultrafast SEF appears as a broad background in SERRS spectra when the SEF rate becomes faster than the vibrational decay rate of the excited electronic states [6,7]. Spectroscopic methods for obtaining absorption, Rayleigh scattering, and ultra-fast SEF from HSs were developed to examine the spectral shapes of the $\hat{\mathbf{A}} \cdot \hat{\mathbf{p}}$ term, $\hat{\mathbf{A}}^2$ term, and EM enhancement, respectively. We found that absorption spectra show EM enhancement more consistently than Rayleigh scattering spectra, revealing that ultrafast SEF can be described by a two-step process using the two $\hat{\mathbf{A}} \cdot \hat{\mathbf{p}}$ terms. This result indicates that the dipole approximation is applicable in describing EM interactions inside HSs. We also showed that spectral inconsistency between Rayleigh scattering and EM enhancement is induced by subradiant plasmon resonance, which spectra are visualized in the absorption not in Rayleigh scattering, with electromagnetic calculation with changes to the morphology of the dimers.

## II. THEORITICAL MODEL

We explained the relationship between the $\hat{\mathbf{A}} \cdot \hat{\mathbf{p}}$ and $\hat{\mathbf{A}}^2$ terms and the linear optical processes i.e., fluorescence, Rayleigh scattering, and spontaneous Raman scattering [15]. The Hamiltonian describing a set of molecular electrons interacting with an EM field are

$$\hat{H} = \hat{H}_e + \hat{H}_{ph} + \hat{H}_{int}, \quad (1)$$

where $\hat{H}_e$ and $\hat{H}_{ph}$ are the free Hamiltonians of the electrons and the EM field, respectively. More correctly, $\hat{H}_e$ expresses a Hamiltonian of the molecular excitons strongly coupled with vacuum fluctuation of the EM field at a HS as shown in Fig.

1(b3) [14]. $\hat{H}_{int}$, which indicates the EM interactions between electrons and the EM field, is described as

$$\hat{H}_{int} = \sum_{i=1}^{N} -\frac{e_i}{m_i}\hat{\mathbf{A}}(\mathbf{r}_i)\cdot\hat{\mathbf{p}}_i + \sum_{i=1}^{N} \frac{e_i^2}{2m_i}\hat{\mathbf{A}}^2(\mathbf{r}_i), \quad (2)$$

where $e_i$, $m_i$, and $\mathbf{r}_i$ are the charge, effective mass, and position of the $i$th electron, respectively [15]. The derivation of Eq. (2) is described in SI. $\hat{\mathbf{p}}_i$ is the operator of the momentum of the $i$th electron as $\hat{\mathbf{p}}_i \equiv \frac{\hbar}{\mathbf{i}}\nabla_i$. $\hat{\mathbf{A}}(\mathbf{r}_i)$ is the operator of the vector potential in $\mathbf{r}_i$ as described as

$$\hat{\mathbf{A}}(\mathbf{r}) = \sqrt{\frac{\hbar}{2\varepsilon_0 V}} \sum_{\mathbf{k}_\gamma} \frac{1}{\sqrt{\omega_\mathbf{k}}} \mathbf{e}_{\mathbf{k}_\gamma} (\hat{a}_{\mathbf{k}_\gamma} e^{i\mathbf{k}\cdot\mathbf{r}_i} + \hat{a}_{\mathbf{k}_\gamma}^\dagger e^{-i\mathbf{k}\cdot\mathbf{r}_i}), \quad (3)$$

where $\hbar$, $\omega_\mathbf{k}$, $\hat{a}_{\mathbf{k}_\gamma}^\dagger$ and $\hat{a}_{\mathbf{k}_\gamma}$ are Planck constant, an angular frequency of light, a creation operator, and an annihilation operator and $\gamma(=1\text{ or }2)$ is direction of polarization. $V$ in Eq. (3) is the mode volume of the EM field. Note that the origin of EM enhancement of plasmonic systems is the small value of $V$ [1]. The first and second terms of Eq. (2) are called the $\hat{\mathbf{A}}\cdot\hat{\mathbf{p}}$ term ($\hat{H}_{int}(AP)$) and the $\hat{\mathbf{A}}^2$ term ($\hat{H}_{int}(A^2)$), respectively. The time-dependent Schrödinger equation for describing interactions between molecular electrons and the EM field in this system is as follows:

$$\hat{H}\Psi(\mathbf{r},t) = i\hbar\frac{\partial\Psi(\mathbf{r},t)}{\partial t}, \quad (4)$$

where $\Psi(\mathbf{r},t)$ is the wave function of the entire system. In the case $\hat{H}_e + \hat{H}_{ph} \gg \hat{H}_{int}$,

one can obtain a solution for Eq. (4) with perturbation theory. The time-independent Schrödinger equation of the stationary states of the system without EM interactions is as follows:

$$\left(\hat{H}_e + \hat{H}_{ph}\right)\Psi_0^n(\mathbf{r}) = E_0^n \Psi_0^n(\mathbf{r}), \quad n = 1, 2, 3, \cdots \quad (5)$$

where $\Psi_0^n(\mathbf{r})$ and $E_0^n$ are the wave function and energy of the system of the $n$th state, respectively, where the subscript 0 denotes that these quantities are associated with the unperturbed system. In perturbation theory, $\Psi(\mathbf{r},t)$ is described as a series of $e^{-iE_0^n t/\hbar}\Psi_0^n(\mathbf{r})$, including time dependence $e^{-iE_0^n t/\hbar}$ as

$$\Psi(\mathbf{r},t) = \sum_n b_n(t) e^{-iE_0^n t/\hbar} \Psi_0^n(\mathbf{r}), \quad (6)$$

where $b_n(t)$ is the amplitude of $\Psi(\mathbf{r},t)$. By assuming that the system initially in $\Psi_0^i$ at $t=0$ starts to be perturbated by $\hat{H}_{int}$, $|b_f(t)|^2$ of the first and second order perturbation terms can be derived as

$$|b_f(t)|^2 \propto \left|\langle\Psi_0^f|\hat{H}_{int}|\Psi_0^i\rangle + \sum_{n \neq i}\langle\Psi_0^f|\hat{H}_{int}|\Psi_0^n\rangle\langle\Psi_0^n|\hat{H}_{int}|\Psi_0^i\rangle\right|^2, \quad (7)$$

where $\langle\Psi_0^f|\hat{H}_{int}|\Psi_0^i\rangle$ and $\sum_{n \neq i}\langle\Psi_0^f|\hat{H}_{int}|\Psi_0^n\rangle\langle\Psi_0^n|\hat{H}_{int}|\Psi_0^i\rangle$ indicate the transitions related to the first and second orders of the perturbation term, respectively, and the subscripts $f$ and $n$ denote the final and intermediate states, respectively.

We discuss the transitions induced by $\langle\Psi_0^f|\hat{H}_{int}|\Psi_0^i\rangle$ in Eq. (7). This matrix element is separated into $\langle \Psi_0^f | \hat{H}_{int}(AP) | \Psi_0^i \rangle$ and $\langle \Psi_0^f | \hat{H}_{int}(A^2) | \Psi_0^i \rangle$ using Eq. (2). Using the dipole approximation of a $\hat{\mathbf{A}} \cdot \hat{\mathbf{p}}$ term as in Eq. (S44), the first term corresponds to a one-photon absorption or one-emission well known as Fermi's golden rule and is described as

$$\langle \Psi_0^f | \hat{H}_{int}(AP) | \Psi_0^i \rangle \propto \sum_{i=1}^{N} \frac{\mathbf{i}}{\hbar} (\varepsilon_f - \varepsilon_i) \langle f | e_i \mathbf{r}_i | i \rangle, \quad (8)$$

where $|i\rangle$ and $|f\rangle$ are the eigenfunctions of the initial and final states of the electron system, respectively, and $\varepsilon_i$ and $\varepsilon_f$ are their energies, respectively, as in Eqs. (S28) and (S29). Fluorescence is composed of a one-photon absorption and emission; thus, the two-step one-photon process of Eq. (8) $\langle f | e_i \mathbf{r}_i | i \rangle \neq 0$ requires $|i\rangle \neq |f\rangle$. Under the dipole approximation of a $\hat{\mathbf{A}}^2$ term as in Eq. (S50), the latter term is described as

$$\langle \Psi_0^f | \hat{H}_{int}(A^2) | \Psi_0^i \rangle \propto \sum_{i=1}^{N} \frac{\mathbf{i}}{\hbar} \langle f \| i \rangle. \quad (9)$$

$\langle f \| i \rangle$ can be nonzero for $|i\rangle = |f\rangle$, meaning that the two-photon process of Eq. (9) corresponds to Rayleigh scattering.

We discuss the transitions using the second order perturbation term $\sum_{n \neq i} \langle \Psi_0^f | \hat{H}_{int} | \Psi_0^n \rangle \langle \Psi_0^n | \hat{H}_{int} | \Psi_0^i \rangle$ in Eq. (7). The matrix elements corresponding to Rayleigh and Raman scattering should include $\hat{H}_{int}(AP)$ as a non-zero term. Thus, this term is rewritten as

$$\sum_{n\neq i}\left\langle\Psi_0^f\left|\hat{H}_{\text{int}}(AP)\right|\Psi_0^n\right\rangle\left\langle\Psi_0^n\left|\hat{H}_{\text{int}}(AP)\right|\Psi_0^i\right\rangle. \quad (10)$$

The matrix element of Eq. (10) can contribute to Rayleigh scattering for $|i\rangle=|f\rangle$ and Raman scattering for $|i\rangle\neq|f\rangle$. The Rayleigh scattering intensity for Eq. (10) may be much weaker than that for Eq. (9) because of the higher order of the perturbation. Thus, the matrix element in Eq. (10) mainly contributes to Raman scattering. In short, Rayleigh scattering is generated by $\left\langle\Psi_0^f\left|\hat{H}_{\text{int}}(A^2)\right|\Psi_0^i\right\rangle$, fluorescence is a two-step process of $\left\langle\Psi_0^f\left|\hat{H}_{\text{int}}(AP)\right|\Psi_0^i\right\rangle$, and Raman scattering is generated by $\sum_{n\neq i}\left\langle\Psi_0^f\left|\hat{H}_{\text{int}}(AP)\right|\Psi_0^n\right\rangle\left\langle\Psi_0^n\left|\hat{H}_{\text{int}}(AP)\right|\Psi_0^i\right\rangle$. Regarding the contribution of $\hat{H}_{\text{int}}(AP)$ to fluorescence and Raman scattering, the spectral shapes of the EM enhancement of ultra-fast SEF and SERRS are expected to be correlated to the absorption spectra of the coupled systems. If the dipole approximation breaks down at HSs, the $\hat{\mathbf{A}}\cdot\hat{\mathbf{p}}$ and $\hat{\mathbf{A}}^2$ terms can contribute to Rayleigh scattering and absorption (or emission or Raman scattering), and therefore such a spectral correlation cannot be observed.

## III. MATERIALS AND METHODS

We describe the synthesis and spectroscopic investigation of the HSs of a coupled system composed of silver NP dimers including dye molecules. We have been studied that the coupling energy of the system reaches several tens to hundreds meV [20-22].

Colloidal silver NPs (mean diameter ~30 nm, $1.10\times10^{-10}$ M) were prepared using the Lee and Meisel method [23]. An equal amount (to the NP dispersion) of an R6G aqueous solution ($1.28\times10^{-8}$ M) was added to the NP dispersion with NaCl (5 mM) and left for 30 min for the aggregation of NPs with including R6G molecules in the HSs. The final concentrations of the R6G solutions ($6.34\times10^{-9}$ M) and NP dispersion ($5.5\times10^{-11}$ M) are almost identical to the reported SM SERRS conditions as shown by the results of the two-analyte or isotope technique [24,25]. Thus, we may detect SERRS and SEF signals exclusively from a small number of molecules inside the HSs of the NP aggregates [26]. The sample solution was dropped onto a slide glass plate coated with Poly-D-Lysine and sandwiched with a cover glass plate. This sample plate was set onto the stage of an inverted optical microscope (IX-71, Olympus, Tokyo). The scattering (extinction) spectra of single NP aggregates were measured by illumination with white light from a 50-W halogen lamp through a dark-field condenser (N.A. 0.92). SERRS and ultrafast SEF spectra of identical aggregates were measured by illumination with an unpolarised polarised excitation green laser beam (2.33 eV (532 nm), 3.5 W/cm$^2$) from a CW Nd3+: YAG laser (DPSS 532, Coherent, Tokyo). Note that the SERRS-active aggregates were almost dimers, if we selected the aggregates showing a clear dipole plasmon resonance with maxima around 1.7–2.1 eV [26].

In measuring the scattering (extinction) spectra, the N.A. of the objective lens (LCPlanFL 100×, Olympus, Tokyo) was set to be 0.6 (1.3) to realize dark- (bright-) field illumination. Gold NPs (mean diameters of 60, 80, and 100 nm; EMGC40, Funakoshi, Japan) were used to convert the scattering (extinction) intensities into their cross-sections [27]. The relationship between the cross-sections of extinction $\sigma_{ext}(\omega)$, Rayleigh scattering $\sigma_{sca}(\omega)$, and absorption $\sigma_{abs}(\omega)$ is $\sigma_{ext}(\omega) = \sigma_{sca}(\omega) + \sigma_{abs}(\omega)$ [28], where $\omega$ is the angular frequency of light. Thus, $\sigma_{abs}(\omega)$ is derived by subtracting $\sigma_{sca}(\omega)$ from $\sigma_{ext}(\omega)$. Figures 2(a1)–1(a3) show the scheme for obtaining $\sigma_{abs}(\omega)$.

## IV. RESULTS AND DISCUSSION

Raman and fluorescence processes are composed of excitation and de-excitation transitions consisting of a two-photon process as shown in Eq. (10) and a two-step one photon process as shown in Eq. (9), respectively. Thus, the EM enhancement factor of SERRS and ultra-fast SEF is described as a product of an excitation enhancement factor $F_R(\omega_{ex})$ and a de-excitation factor $F_R(\omega)$ due to plasmon resonance as

$$F_R(\omega_{ex}, \mathbf{r}) F_R(\omega, \mathbf{r}) = \left| \frac{E_{loc}(\omega_{ex}, \mathbf{r})}{E_I(\omega_{ex})} \right|^2 \times \left| \frac{E_{loc}(\omega, \mathbf{r})}{E_I(\omega)} \right|^2, \quad (11)$$

where $E_I$ and $E_{loc}$ indicate the amplitudes of the incident and enhanced local electric fields, respectively; $\omega_{ex}$ and $\omega$ denote the angular frequencies of the incident and Raman-scattered light, respectively; and **r** is the position of a molecule in a HS [29]. The mode volume of the EM field confined to inside a HS as in Eq. (3) mainly determines the value of $F_R$ [1]. The relationship between Eq. (11) and the mode volume in Eq. (3) is explained as the effective Purcell's factor in Ref. 1. Equation (11) indicates the cross-section of ultra-fast SEF $\sigma_{uS}(\omega)$, as follows:

$$\sigma_{uS}(\omega_{ex},\omega) = F_R(\omega_{ex},\mathbf{r}) F_R(\omega,\mathbf{r}) \sigma_F(\omega_{ex},\omega), \quad (12)$$

where $\sigma_F(\omega)$ is the cross-section of ultra-fast fluorescence without EM enhancement. The spectral shape of $\sigma_F(\omega)$ is obtained by averaging $\sigma_{uS}(\omega)$ spectra from a large number of NP aggregates. In other words, the spectral shape of $F_R(\omega)$ in Eq. (12) is flattened by the averaging effect [30]. Thus, $F_R(\omega)$ can be obtained by dividing $\sigma_{uS}(\omega)$ by $\sigma_F(\omega)$. Figures 2(b1)–1(b3) show the scheme of division for obtaining $F_R(\omega)$.

Figures 3(a1)–3(a4) exhibit spectra of $\sigma_{sca}(\omega)$, $\sigma_{abs}(\omega)$ and $F_R(\omega)$ for four dimers showing single Lorentzian maxima, which are dipole-dipole (DD) coupled plasmon resonances [26] between approximately 1.9 to 2.2 eV. One may notice that the spectral shapes of $F_R(\omega)$ look more consistent with those of $\sigma_{abs}(\omega)$ than with those of $\sigma_{sca}(\omega)$. Figure 3(a5) shows the relationship between the peak energies of $\sigma_{sca}(\omega)$ $\hbar\omega_{sca}$, those of $\sigma_{abs}(\omega)$ $\hbar\omega_{abs}$ against those of $F_R(\omega)$ $\hbar\omega_R$. The positions of $\hbar\omega_{abs}$ overlap substantially with those of $\hbar\omega_R$. This result indicates that the origin of $F_R(\omega)$ is $\hat{H}_{int}(AP)$ in Eq. (8),

which shows that the dipole approximation is applicable for the HSs of the present coupled system. Regarding the size of R6G molecule around 1 nm and that of a HS of several nm, the applicability of dipole approximation is rather surprising. We consider that the molecules exist on the saddle point of the electromagnetic potential inside the HS. The positions of $\hbar\omega_{sca}$ are always redshifted from those of $\hbar\omega_R$. These redshifts may be induced by the contribution of a subradiant plasmon e.g., dipole-quadrupole (DQ) coupled plasmon, whose resonances appear in the higher energy region of $\hbar\omega_{sca}$, to $F_R(\omega)$ [31].

Figures 3(b1)–3(b4) show spectra of $\sigma_{sca}(\omega)$, $\sigma_{abs}(\omega)$ and $F_R(\omega)$ for the four dimers with more complicated spectral shapes than those in Figs. 3(a1)–3(a4). The $\sigma_{abs}(\omega)$ and $F_R(\omega)$ spectra show several common spectral maxima which do not exist in $\sigma_{sca}(\omega)$, as indicated by the black arrows. These common maxima reveal that the spectral shapes of $F_R(\omega)$ are determined by $\sigma_{abs}(\omega)$. Figure 3(b5) shows the relationship between $\hbar\omega_{sca}$, $\hbar\omega_{abs}$ against $\hbar\omega_R$. For $\hbar\omega_{sca}$, we selected the lowest energy spectral maxima, which corresponds to the superradiant plasmon resonance. The positions of $\hbar\omega_{abs}$ always agree with those of $\hbar\omega_R$. From the correlation between $\hbar\omega_{abs}$ and $\hbar\omega_R$, we conclude that the origin of $F_R(\omega)$ is $\hat{H}_{int}(AP)$ in Eq. (8).

The results of Fig. 3 indicate that the contribution of $\hat{H}_{int}(AP)$ to $F_R(\omega)$ at HSs.

These results can be reproduced by electromagnetism in the form of a spectral correlation between $\sigma_{abs}(\omega)$ and $F_R(\omega)$. Thus, we examined this correlation using the FDTD method (EEM-FDM Ver.5.1, EEM Co., Ltd., Japan). The complex refractive index of silver NPs was taken from Ref. 32. The detailed calculation conditions for reproducing the experimental conditions have been described elsewhere [26]. Figures 4(a) and 4(b) show the typical spectra of $\sigma_{sca}(\omega)$ and SERRS with ultra-fast SEF of the dimers with their SEM images (JSM-6700F, JEOL). The symmetric dimers exhibit a good spectral correlation between $\sigma_{sca}(\omega)$ and SERRS with ultra-fast SEF as in Fig. 4(a). In contrast, the asymmetric dimers do not exhibit such correlation. Thus, the experimentally obtained spectral relationships between $\sigma_{sca}(\omega)$, $\sigma_{abs}(\omega)$ and $F_R(\omega)$ were examined by changing the degree of asymmetry of the dimers.

Figure 4(c) illustrates the light excitation and a dimer composed of two spherical NPs with diameters $D_1$ and $D_2$ while maintaining a gap of 1 nm for FDTD calculation. This value was selected to ensure consistency between experiments and calculations [22,31]. We consider that a small number of R6G molecules, with a size of around 1.0 nm, were inserted into the gap and determined the gap distance [31]. The spectra of $\sigma_{sca}(\omega)$, $\sigma_{abs}(\omega)$, and $F_R(\omega)$ at the gap with a phase retardation of $E_{loc}$ against $E_I$ were calculated by changing the ratio $D_1/D_2$. The excitation polarization direction was set to be parallel to the long axis of the dimer because intense $F_R(\omega)$ is generated along this direction at a HS. Figures 4(d1) and 4(d2) illustrate the charge distributions of DD and DQ coupled plasmons, respectively, by changing $D_1/D_2$. In our calculations, the non-local effect, which reduces $F_R(\omega)$ by Landau damping due to unscreened surface electrons [8], was not considered because Landau damping does not change the spectral shape of $F_R(\omega)$ but rather its intensity [8].

Figures 5(a1)–5(a3) show $\sigma_{sca}(\omega)$ and $\sigma_{abs}(\omega)$ by changing the ratio $D_1/D_2$ while keeping the value of $D_1$ as 30 nm. As the degree of asymmetry is increased, the shapes of $\sigma_{sca}(\omega)$ and $\sigma_{abs}(\omega)$ deviate from each other. Figure 5(a4) shows the relationship between $\hbar\omega_{sca}$, $\hbar\omega_{abs}$ and $D_2$. The higher energy shifts in $\sigma_{abs}(\omega)$ are clearly observed for $D_2 > 80$ nm, indicating that the quadrupole of NP for $D_2$ resonantly interacts with the dipole of another NP as in Fig. 4(d2). Figures 5(b1)–5(b3) show $\sigma_{sca}(\omega)$ and $F_R(\omega)$ by changing the ratio $D_1/D_2$. As the degree of asymmetry is increased, the shape of $F_R(\omega)$ exhibits common spectral deviation from that of $\sigma_{abs}(\omega)$, as in Figs. 5(a1)–5(a3). Figure 5(b4) shows the relationship between $\hbar\omega_{sca}$, $\hbar\omega_F$ and $D_2$ for the positions of phase retardations of $\phi = 90°$ (DD coupled plasmon resonating with incident light) and $180°$ (DQ coupled plasmon resonating with the DD coupled plasmon). The position of $\phi = 90°$ agrees with that of $\hbar\omega_{sca}$, indicating a DD coupled plasmon. The position of $\hbar\omega_F$

follows that of $\phi = 90°$ for $D_2$ between 30 and 80 nm. Then, it moves to $\phi = 180°$, indicating that the origin of $F_R(\omega)$ switches from a DD to a DQ coupled resonance with an increase in the degree of asymmetry. The switch is caused by near-field interactions between the DD and DQ coupled resonance. Figures 5(c1)–5(c3) show $\sigma_{abs}(\omega)$ and $F_R(\omega)$ for changing ratios of $D_1/D_2$. As expected from Figs. 5(a) and 5(b), the maxima of $F_R(\omega)$ are similar to those of $\sigma_{abs}(\omega)$. The positions of $\hbar\omega_F$ also exhibit identical behaviour to those of $\hbar\omega_{abs}$, supporting the experiments that both $\sigma_{abs}(\omega)$ and $F_R(\omega)$ are generated by the $\hat{\mathbf{A}}\cdot\hat{\mathbf{p}}$ term. The spectral deviation between $\sigma_{abs}(\omega)$ and $F_R(\omega)$ may be due to a contribution from light absorption outside the HS. These calculations of electromagnetism support the experimentally obtained conclusion that the origin of $F_R(\omega)$ is $\hat{H}_{int}(AP)$.

## V. CONCLUTION

In this study, we investigated the origin of EM enhancement inside HSs of coupled systems between plasmon polariton and molecular exciton. The EM interaction between a photon and an exciton is expressed by the $\hat{\mathbf{A}}\cdot\hat{\mathbf{p}}$ and $\hat{\mathbf{A}}^2$ terms. Thus, we developed an experimental method for evaluating these terms and EM enhancement using absorption, Rayleigh scattering, and ultrafast SEF of the single coupled systems, respectively. A spectral comparison revealed that the absorption spectra can correctly reproduce the EM enhancement. This result indicates that ultrafast SEF is a two-step process with the $\hat{\mathbf{A}} \cdot \hat{\mathbf{p}}$ terms, showing the dipole approximation is applicable to the present HSs. The subradiant resonance, which induces spectral deviation between the scattering and absorption, was observed in the absorption and EM enhancement spectra. These results were well supported by calculations based on electromagnetism. This method will be applied to EM enhancement of various cQED systems and plasmonic spectroscopies [34-38].

## ACKNOWLEDGMENTS

This work was supported by the JSPS KAKENHI Grant-in-Aid for Scientific Research (C), number 21K04935.

**Figure captions**

FIG. 1. (a1) and (b1) Schematic images of a molecule in a free space and in a HS between two silver NP, respectively, excited by incident light. (a2) and (b2) The intensity distribution of incident light around a molecule in a free space and in a HS, respectively. (a3) and (b3) Energy level diagrams of a molecule in a free space and in a HS, respectively. Here, $|g\rangle$ and $|e\rangle$ are the ground and excited states of the molecule assumed to be a two-level system, respectively; $|0\rangle$ and $|1\rangle$ are the zero-photon and one-photon states of the plasmon resonance; $2\hbar g$ indicates vacuum Rabi splitting, where $\hbar g$ corresponds to the coupling energy between the two-level system and plasmon resonance.

FIG. 2. (a) Schematic images of $\sigma_{\text{ext}}(\omega) - \sigma_{\text{sca}}(\omega) = \sigma_{\text{abs}}(\omega)$. (a1)–(a3) Spectra of $\sigma_{\text{ext}}(\omega)$, $\sigma_{\text{sca}}(\omega)$, and $\sigma_{\text{abs}}(\omega)$ of single silver NP dimer, respectively. Insets of (a1) and (a2) are the extinction and scattering images observed under a bright- and dark-field microscope, respectively. The positions of the peaks in $\sigma_{\text{sca}}(\omega)$ at $\hbar\omega_{\text{sca}}$ and $\sigma_{\text{abs}}(\omega)$ at $\hbar\omega_{\text{abs}}$ are indicated in the panels. (b) Schematic images of $\sigma_{\text{S}}(\omega_{\text{ex}},\omega_{\text{em}})/\sigma_{\text{F}}(\omega_{\text{ex}},\omega_{\text{em}}) = F_{\text{R}}(\omega_{\text{ex}})F_{\text{R}}(\omega_{\text{em}}) \propto F_{\text{R}}(\omega_{\text{em}})$. (b1) and (b2) Spectra of SERRS with ultra-fast SEF of single silver NP dimer and large silver NP aggregate, respectively. Insets of (b1) and (b2) show the SERRS with ultra-fast SEF images of them. (b3) Spectrum of $F_{\text{R}}(\omega)$. The position of the peak of $F_{\text{R}}(\omega)$ at $\hbar\omega_{\text{R}}$ is indicated in the panel.

FIG. 3. (a1)–(a4) Spectra of $\sigma_{\text{sca}}(\omega)$ (blue lines), $\sigma_{\text{abs}}(\omega)$ (green lines), and $F_{\text{R}}(\omega)$ (red lines) of single dimers exhibiting single Lorentzian maxima, respectively. (a5) Relationships between $\hbar\omega_{\text{sca}}$ (blue open circles), $\hbar\omega_{\text{abs}}$ (green open circles) and $\hbar\omega_{\text{R}}$.

(b1)–(b4) Spectra of $\sigma_{sca}(\omega)$ (blue lines), $\sigma_{abs}(\omega)$ (green lines), and $F_R(\omega)$ (red lines) of single dimers, respectively, showing more complicated spectral shapes than those in (a1)–(a4). (b5) Relationships between $\hbar\omega_{sca}$ (blue open circles), $\hbar\omega_{abs}$ (green open circles) and $\hbar\omega_R$.

FIG. 4. (a) Ultra-fast SEF with SERRS (red line) and $\sigma_{sca}(\omega)$ spectrum (blue line) of symmetric dimer. Insets are SEM images of the dimer. (b) Ultra-fast SEF with SERRS (red line) and $\sigma_{sca}(\omega)$ spectrum (blue line) of asymmetric dimer. Insets are SEM images of the dimer. Scale bars are 100 nm. (c) FDTD calculation setup for $\sigma_{sca}(\omega)$, $\sigma_{abs}(\omega)$, and $F_R(\omega)$ of a single NP dimer composed of two NP with diameters with $D_1$ and $D_2$. The gap was set to 1 nm. The position of the HS is indicated by a red dot. The excitation polarization direction is parallel to the long axis of the dimer. (d1) and (d2) Schematic images of the charge distributions of DD and DQ coupled plasmons of the symmetric and asymmetric dimers, respectively.

FIG. 5. (a1)–(a3) $D_2$ dependence of the $\sigma_{sca}(\omega)$ and $\sigma_{abs}(\omega)$ spectra of dimers with $D_1$ of 30 nm and $D_2$ of 30, 80, and 120 nm, respectively. (a4) $D_2$ dependences of $\hbar\omega_{sca}$ (blue open circles) and $\hbar\omega_{abs}$ (green open circles) for dimers with $D_1$ of 30 nm. (b1)–(b3) $D_2$

dependence of the $\sigma_{sca}(\omega)$ and $F_R(\omega)$ spectra with $D_1$ of 30 nm and $D_2$ of 30, 80, and 120 nm, respectively. (b4) $D_2$ dependences of $\hbar\omega_{sca}$ (blue open circles), $\hbar\omega_F$ (red open circles), $\phi = 90°$ (black solid line, DD resonance), and $\phi = 180°$ (black dashed line, DQ resonance) for dimers with $D_2$ of 30 nm. (c1)–(c3) $D_2$ dependence of the $\sigma_{abs}(\omega)$ and $F_R(\omega)$ spectra with $D_1$ of 30 nm and $D_2$ of 30, 80, and 120 nm, respectively. (c4) $D_2$ dependences of $\hbar\omega_{abs}$ (blue open circles), $\hbar\omega_F$ (red open circles), $\phi = 90°$ (black solid line, DD resonance), and $\phi = 180°$ (black dashed line, DQ resonance) for dimers with $D_2$ of 30 nm.

Figure 1

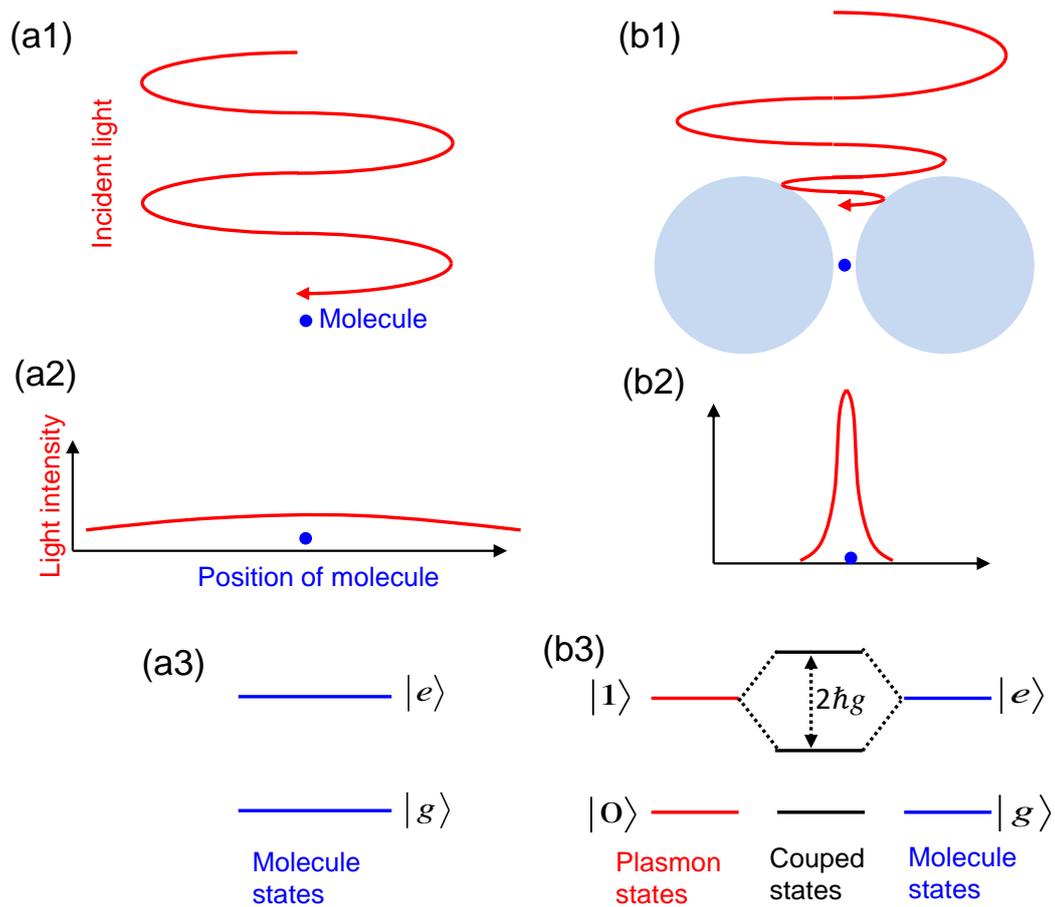

Figure 2

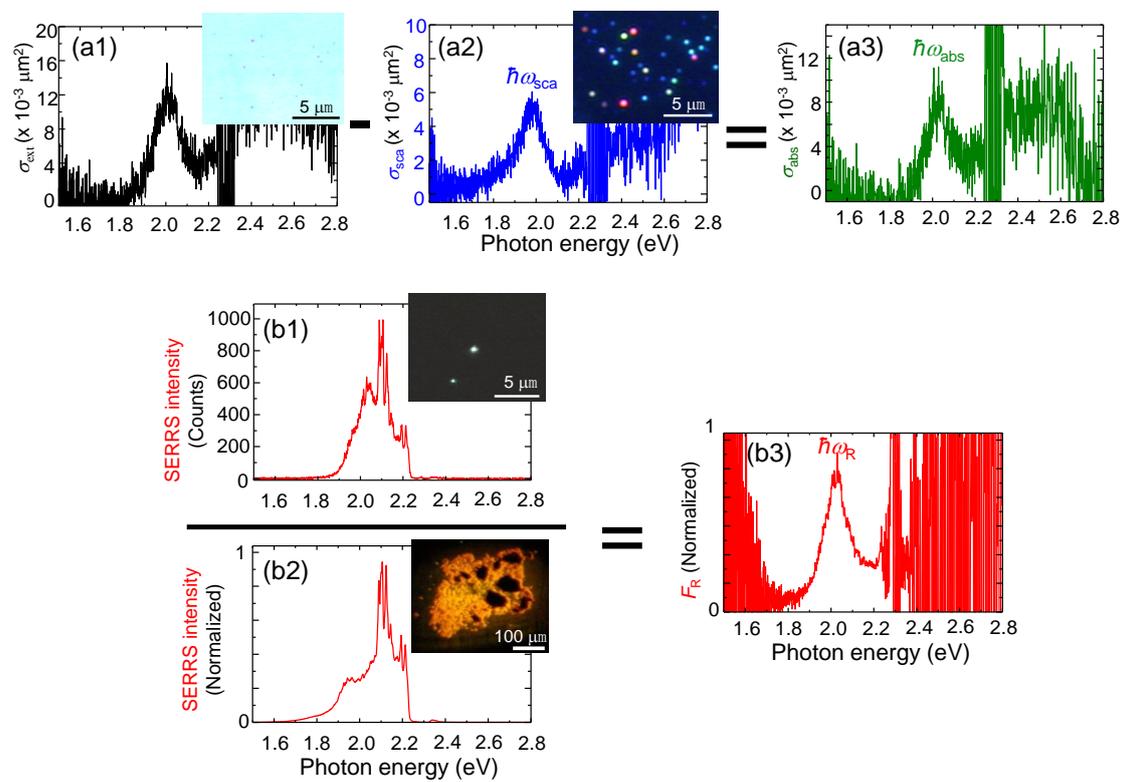

Figure 3

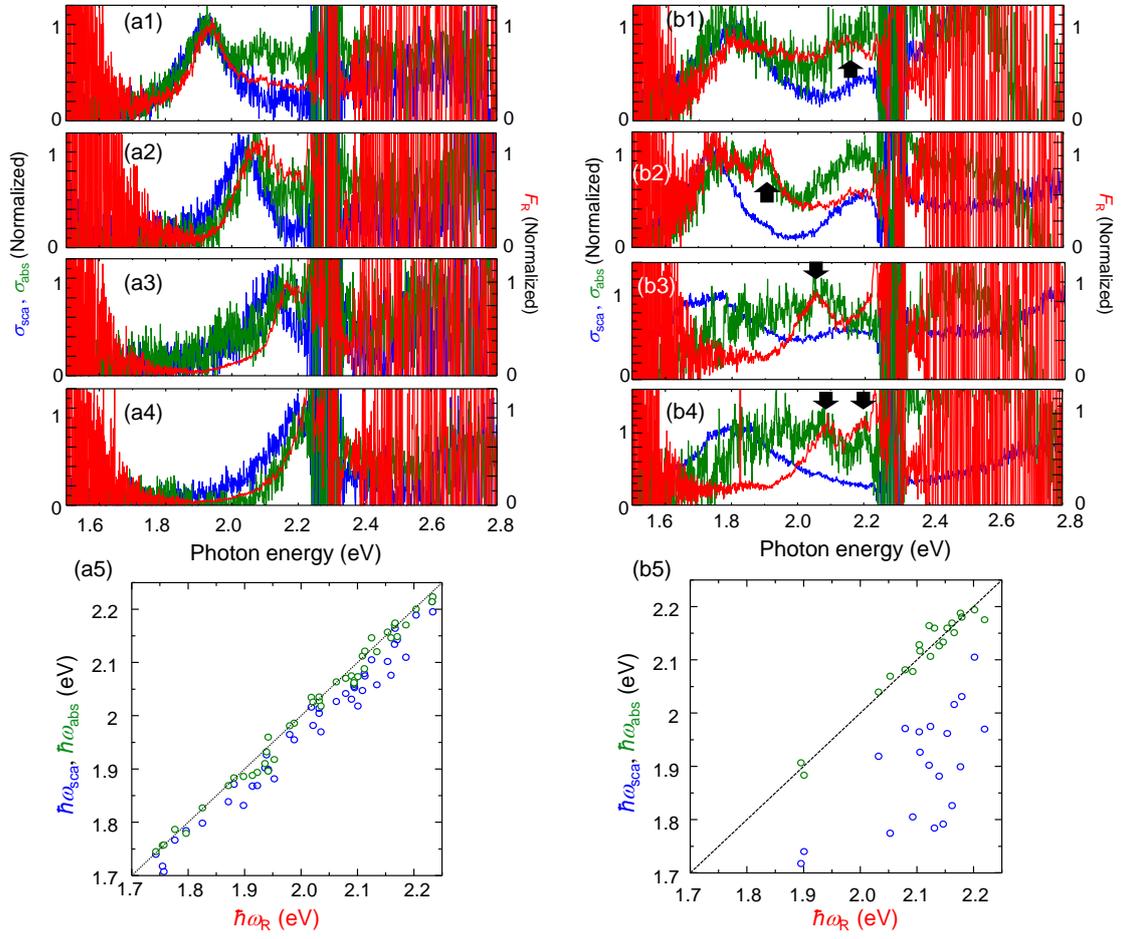

Figure 4

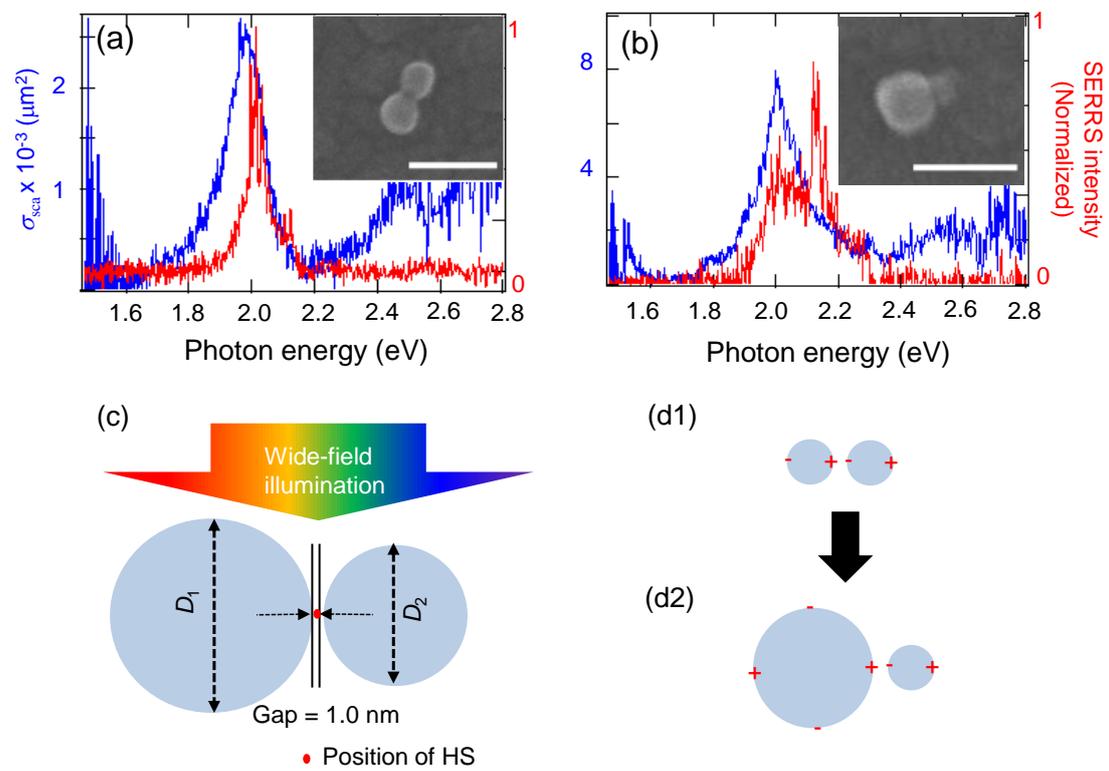

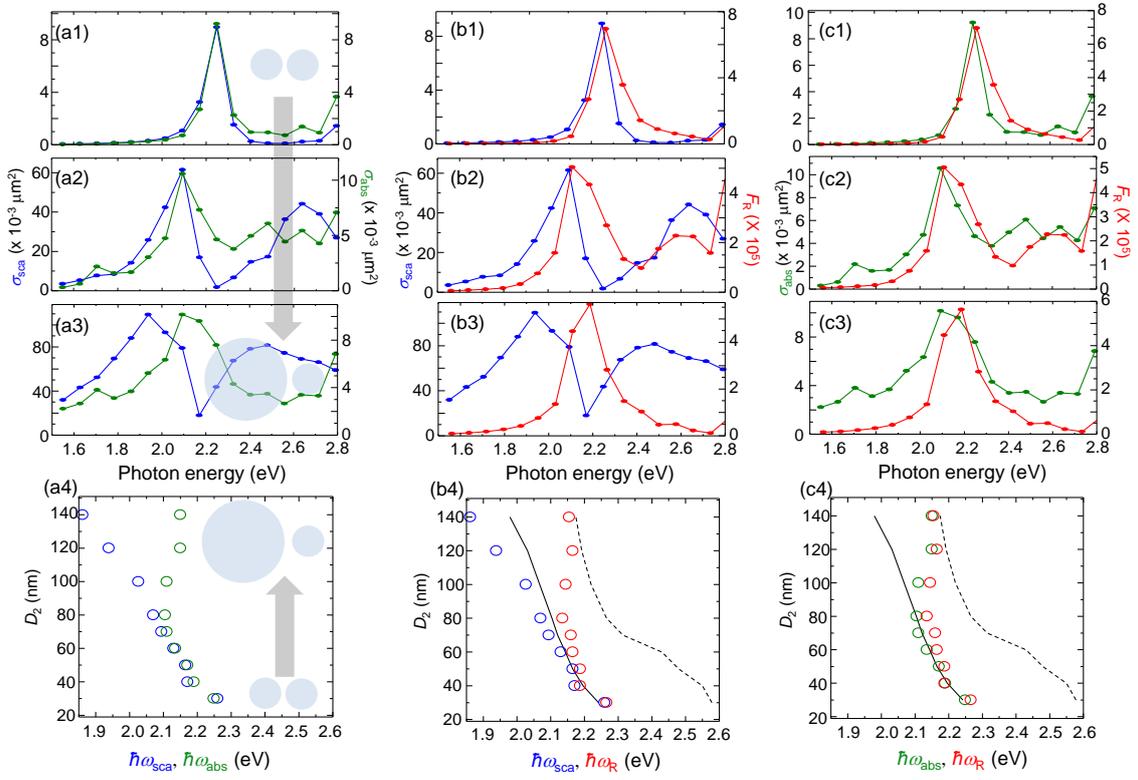